\documentclass[aps,prl,twocolumn,superscriptaddress]{revtex4}
\usepackage{float}
\usepackage{graphicx}
\usepackage{dcolumn}
\usepackage{bm}
\usepackage{color}
\usepackage{amsmath}
\usepackage{amssymb}
\usepackage{xcolor}
\usepackage{mathdots}
\newcommand{\pdag}{{\phantom{\dagger}}}

\newcommand{\eph}{{\it e}-ph}

\newcommand{\D}{\mathrm{d}}
\allowdisplaybreaks

\begin{document}
\title{Accelerating lattice quantum  Monte Carlo simulations using 
artificial neural networks: an application to the Holstein model}
\author{Shaozhi Li}
\affiliation{Department of Physics and Astronomy, The University of Tennessee, Knoxville, Tennessee 37996, USA}
\affiliation{Department of Physics, University of Michigan,  Ann Arbor, Michigan 48109, USA}
\author{Philip M. Dee}
\affiliation{Department of Physics and Astronomy, The University of Tennessee, Knoxville, Tennessee 37996, USA}
\author{Ehsan Khatami}
\affiliation{Department of Physics and Astronomy, San Jos\'{e} State University, San Jos\'{e}, California 95192, USA}
\author{Steven Johnston}
\affiliation{Department of Physics and Astronomy, The University of Tennessee, Knoxville, Tennessee 37996, USA}
\affiliation{Joint Institute for Advanced Materials at the University of Tennessee, Knoxville, Tennessee 37996, USA}
\date{\today}

\begin{abstract}
Monte Carlo (MC) simulations are essential computational approaches with
widespread use throughout all areas of science. We present a method for
accelerating lattice MC simulations using fully-connected and convolutional 
artificial neural networks that are trained to perform {\it local} and
{\it global} moves in configuration space, respectively. Both networks take local spacetime
MC configurations as input features and can, therefore, be trained using
samples generated by conventional MC runs on smaller lattices 
before being utilized for simulations on larger
systems. This new approach is benchmarked for the case of determinant quantum Monte
Carlo (DQMC) studies of the two-dimensional Holstein model. 
We find that both artificial neural networks are capable of learning an unspecified effective model 
that accurately reproduces the MC configuration weights of the 
original Hamiltonian and achieve an order of magnitude speedup over the conventional 
DQMC algorithm. Our approach is broadly applicable to many classical and 
quantum lattice MC algorithms. 
\end{abstract}

\maketitle
{\it Introduction} --- 
As their full potential becomes apparent, machine learning algorithms are assuming more 
prominent roles in the process of scientific discovery. 
Meanwhile, the boundary lines between 
industry applications of machine learning, data and computer science, and other
disciplines have blurred. Applications ranging from the high-quality feature extraction
from astrophysical images of galaxies~\cite{Schawinski2017MNRAS} to helping
with the real-time data analysis of particle
accelerators~\cite{Edelen2016IEEE,Radovic2018Nature,Scheinker2018PRL}  
at the Fermilab~\cite{Edelen2016IEEE} and the Large Hadron
Collider~\cite{Radovic2018Nature} to discovering phases of
matter~\cite{Carrasquilla2017NatPhys,Wang2016PRB,Zhang2018arXiv} have emerged. 

A series of early studies have underscored the potential for machine learning
in the context of condensed matter physics by using artificial neural networks
and dimension-reduction techniques to locate phase transitions~\cite{Carrasquilla2017NatPhys,Wang2016PRB}, or
represent ground states of quantum many-body systems~\cite{Carleo2017Science}.
Machine learning algorithms have also been employed to help
gain more insight into classical and quantum
systems~\cite{Torlai2016,k_chng_17,d_deng_16,e_vanNieuwenburg_17,y_zhang_17a,
s_wetzel_17,w_hu_17,g_torlai_17,f_schindler_17}
as well as accelerate specific numerical algorithms~\cite{l_huang_17,LiuPRB2017, 
s_li_18, r_fournier_18, j_nelson_18,h_suwa_18}. 
These applications are not only helping to automate and streamline scientific processes  
that could take many years to accomplish with more 
conventional computational approaches, but they are also uncovering 
previously inaccessible phenomena. 

One machine learning application that has attracted significant attention 
is in accelerating MC simulations~\cite{l_huang_17,LiuPRB2,
LiuPRB2017,XuPRB2017,Nagai2017PRB,ShenPRB2018,Chen2018PRB,Nagai2018preprint,e_inack_18,t_song_19}.
For example, in the so-called self-learning Monte Carlo (SLMC) method~\cite{LiuPRB2}, an effective 
bosonic model is trained to mimic 
the statistics of the original Hamiltonian. Once trained, 
the effective model is then used  
to perform the same simulations much more efficiently. The primary advantage of this approach 
is that the action of the effective model is often much easier to compute 
than the action for the full fermion model, thus granting access to larger system 
sizes. This approach has also been 
extended to include correlations in both the real space and imaginary time
domains~\cite{XuPRB2017,Nagai2017PRB}. Despite their power, however, the SLMC methods require    
that the form of the effective model be known {\it a priori}. This limitation can 
be significant as different effective models may be required for the same fermionic model 
as the model parameters, system size, or simulation temperature changes, 
and the overall effectiveness of these approaches is severely limited if the wrong 
effective model is chosen. To overcome this problem, several groups have used artificial neural networks 
to learn the form of the model in some instances~\cite{ShenPRB2018,Nagai2018preprint,e_inack_18,t_song_19} 
({\it e.g.} QMC simulations of the Anderson impurity model); however, generalizing 
this approach to lattice QMC problems has not yet been achieved. One reason for this is 
the fact that such problems typically involve thousands of auxiliary spacetime fields and any 
neural network using that many input features will often generalize poorly.    

\begin{figure*}[t]
\center{\includegraphics[width=\textwidth]{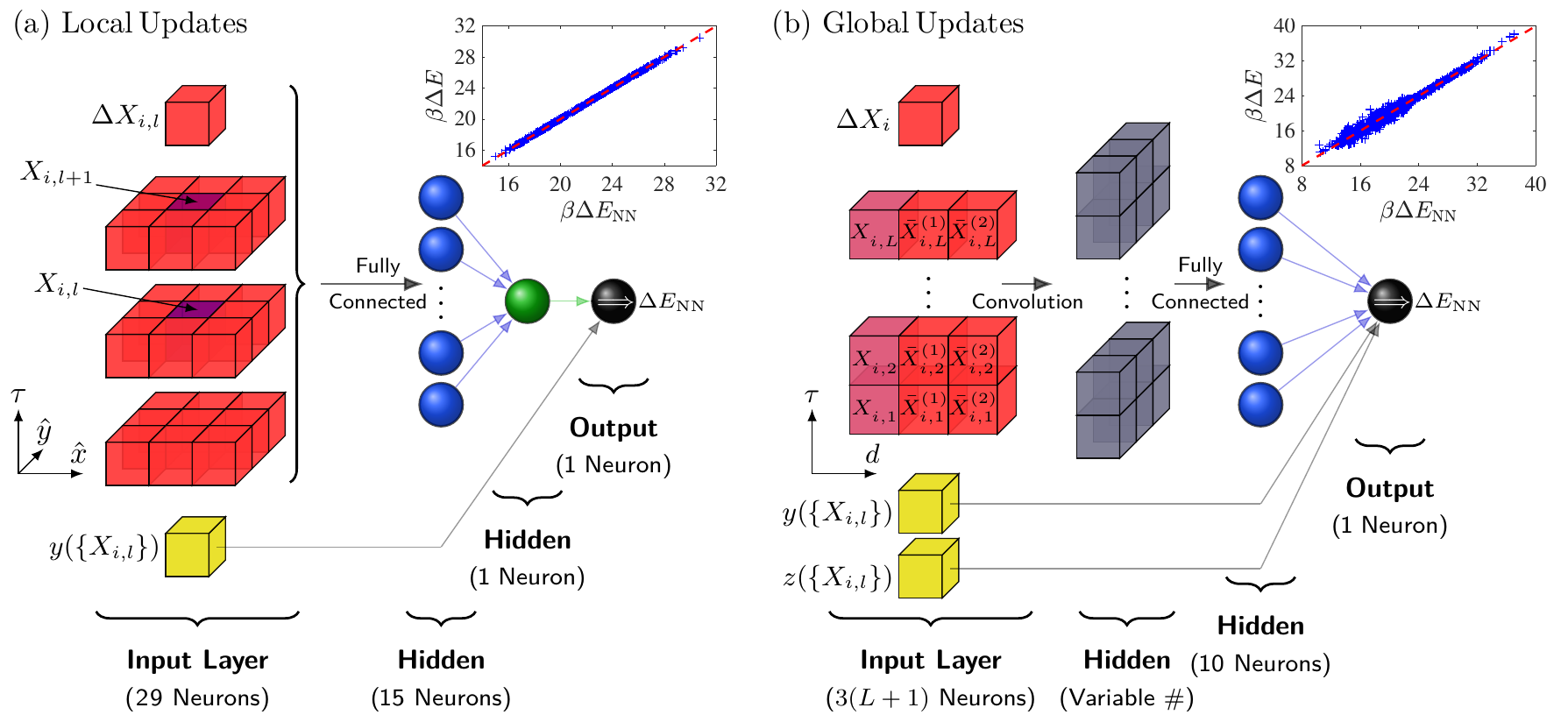}}
\caption{A sketch of the architecture of the (a) fully-connected and (b) convolutional neural networks (CNN)
used to perform local and global updates of the auxilary fields, respectively. 
The first and second hidden layers of the fully-connected neural 
network use softplus activation functions $f(x) = \ln(1+e^x)$, while the output layer 
uses a linear activation function. 
The first and second hidden layers of the CNN 
use sigmoid functions $f(x) = (1+e^{-x})^{-1}$, while the output layer uses a linear function. 
The number of neurons in the first hidden layer of the CNN is 
set by the stride and kernel.  
A measure of the performance of the two networks is presented in the insets,  
which compare the predicted $\beta \Delta E_\text{NN}$ against the exact $\beta \Delta E$ for the 
fully-connected and the CNN, respectively. 
These results were obtained using networks trained on an $N = 6\times 6$ cluster, 
an inverse temperature $\beta=4.1/t$, filling $\langle \hat{n}\rangle = 1$, $\lambda=t/2$, 
and $\Omega=t/2$. 
}
\label{Fig:NN}
\end{figure*}

Here, we show how to 
design artificial neural networks that can be trained to represent an effective bosonic
model for lattice QMC simulations. 
Inspired by applications of the traveling cluster 
approximation to spin-fermion models \cite{TCA,PTCA}, we design fully-connected and 
convolutional neural networks that only require information from surrounding auxiliary fields (see. Fig. \ref{Fig:NN}) 
to perform both {\it local} and {\it global} moves of the MC configurations. 
This method does not suffer from the scaling issues restricting 
other self-learning methods and can be easily generalized across models and parameter 
regimes without changes in the underlying algorithm, provided the neural networks are versatile enough to learn the effective models. As such, this approach can be integrated 
directly into existing MC codes.  
To demonstrate the efficiency of this approach, here we apply it to determinant 
quantum Monte Carlo (DQMC) simulations of the two-dimensional Holstein model. This problem is 
particularly challenging owing to long autocorrelation times \cite{Hohenadler}, 
the need for both local and global MC moves to ensure 
ergodicity \cite{ScalettarPRB1991,JohnstonPRB2013}, and competition between 
charge-density-wave (CDW) and superconducting ground states \cite{DeePRB2019} that 
may require different effective boson models. Reproducing known
results, we obtain an order of magnitude of speedup with our algorithm. 

{\it Model} --- 
The single-band Holstein Hamiltonian \cite{Holstein} is $H=H_0+H_\text{lat}+H_\text{e-ph}$, where
$H_0=-t\sum_{\langle i, j \rangle, \sigma}c_{i,\sigma}^{\dagger}c_{j,\sigma}^{\phantom\dagger}-\mu\sum_{i,\sigma}\hat{n}_{i,\sigma}$, $H_\text{lat}=\sum_{i}\left(\frac{1}{2M}\hat{P}_i^2 + \frac{M\Omega^2}{2}\hat{X}_i^2\right)$ , and $H_\text{e-ph}=g\sum_{i,\sigma}\hat{X}_{i}\hat{n}_{i,\sigma}$.
Here, $\langle \cdots \rangle$ denotes a summation over nearest neighbors;
$c_{i,\sigma}^{\dagger}$ ($c_{i,\sigma}^{\phantom\dagger}$) creates
(annihilates) an electron with spin $\sigma$ on site $i$;
$\hat{n}_{i,\sigma}=c_{i,\sigma}^{\dagger}c^\pdag_{i,\sigma}$ is the particle number
operator; $t$ is the nearest-neighbor hoping integral; $M$ is the ion mass and
$\Omega$ is the phonon frequency; $\hat{X}_i$ and $\hat{P}_i$ are the lattice
position and momentum operators, respectively; and $g$ is the strength of the
{\eph} coupling. Throughout, we set $M=t=1$ as the unit of mass and energy, and 
we study this Hamiltonian on an $N=N_x^2$ square lattice, where $N_x$ is the linear 
size of the cluster.  
To facilitate a direct comparison with a recent state-of-the-art simulation \cite{Chen2018PRB}, 
we focus on $\Omega=t/2$ and dimensionless {\eph} coupling strength  
$\lambda=\frac{g^2}{8t\Omega^2}=0.5$. 

{\it Determinant quantum Monte Carlo} --- DQMC is an auxiliary field, imaginary time technique that computes expectation
values of an observable within the grand canonical ensemble. 
In a DQMC simulation, the imaginary time interval $\tau \in [0,\beta]$ is evenly divided into $L$ 
discrete slices of length $\Delta\tau=\beta/L$ ($= 0.1$ in this work). 
Using the Trotter approximation, the partition function is then given by $Z=\mathrm{Tr}\left(
e^{-\Delta\tau L H} \right)\approx \mathrm{Tr} \left( e^{-\Delta\tau
H_{\text{\eph}}} e^{-\Delta\tau (H_0 + H_\mathrm{lat})}  \right)^L$. 
After integrating out the electronic degrees of freedom, the partition function 
can be reduced to $Z=\int W\left(\{X\}\right) \D X$, where 
the configuration weight is $W\left(\{X\}\right) = 
e^{-S_\mathrm{ph}\Delta \tau} \mathrm{det}
M^{\uparrow} \mathrm{det} M^{\downarrow}$. Here, $\int \D X$ is shorthand for
integrating over all of continuous displacements $X_{i,l}$ defined at 
each spacetime point $(i,l)$, the matrices $M^{\sigma}$ are defined as 
$M^{\sigma}=I+B_L^{\sigma}B_{L-1}^{\sigma}\cdots B_1^{\sigma}$, where $I$ is an
$N\times N$ identity matrix, and $B_{l}^\sigma=e^{-\Delta\tau
H_{\text{\eph}}}e^{-\Delta\tau H_0}$, and
$S_\mathrm{ph}=\frac{M}{2\Delta\tau^2}\sum_{i,l}\left(X_{i,l+1}-X_{i,l}\right)^2 
+ \frac{M\Omega^2}{2}\sum_{i,l}X_{i,l}^{2}$
is lattice's contribution to the total action. Note that $B^\sigma_l$ matrices for the Holstein model
do not depend on spin but are implicitly dependent on the 
fields $X_{i,l}$ through $H_{\text{\eph}}$.
For more details, we refer the reader to Refs. 
\onlinecite{WhitePRB1989,Scalettar1989,Johnston2013PRB}. 
 
As mentioned, two types of MC updates are needed in the simulation.  
The first are local updates of the type $X_{i,l} \rightarrow X^\prime_{i,l}=X_{i,l}+\Delta X_{i,l}$, 
which are made at each spacetime point. The second are global or block updates, 
where the field for a given site are updated 
simultaneously at all timeslices $X_{i,l}\rightarrow X_{i,l}+\Delta X_i,~\forall l\in[0,L]$. 
Such block updates are needed to help move phonon 
configurations out of local minima at low temperatures and large couplings~
\cite{ScalettarPRB1991,JohnstonPRB2013}. DQMC accepts both 
kinds of moves with a probability $p = W\left(\{X^\prime\}\right)/W\left(\{X\}\right) \equiv e^{-\beta \Delta E}$, 
which requires the costly evaluation of matrix determinants. 
Moreover, since the matrices $M^{\sigma}$ depend on the fields, these 
must also be updated after every accepted change in the phonon fields. 
While an efficient update algorithm exists for performing local updates \cite{WhitePRB1989}, 
no such algorithm is known for block updates. The computational cost for performing a full 
sweep of (fast) local updates and block updates is $\mathcal{O}(N^3L)$ and $\mathcal{O}(N^4L)$ \cite{LeeIEEE}, respectively.

To reduce this cost, we train our networks to predict  
$\beta\Delta E$ appearing in the definition of the configuration weight  
given only changes in, and local information of, the phonon fields and their expected  
behavior at large displacements as input. This reduces the total computational 
complexity of determining whether both kinds of updates will be accepted to the time needed 
to evaluate the networks, which is $\mathcal{O}(1)$ for the case of the simply connected network 
and $\mathcal{O}(L)$ for the convolutional neural network. 
As with other SLMC methods, we then use the neural networks to propose many MC updates that are ultimately  
accepted or rejected based on the configuration weights of the original model. 
While determining this final acceptance probability requires the evaluation of the matrix 
determinants, this task can be done  
infrequently enough that a considerable speedup is achieved. 
Another advantage of our approach is that the networks can be trained 
using data generated by the conventional DQMC algorithm 
on inexpensive small clusters before being generalized to 
larger systems. In this way, our method combines the flexibility of neural networks 
with the inexpensive training costs seen in SLMC approaches making use of 
largely local effective models.  

{\it Local Updates} --- Local updates are performed using a fully-connected 
network with two hidden layers [Fig.~\ref{Fig:NN}(a)]. Assuming that the update 
is proposed at spacetime site $(i,l)$, the learning objective is to predict 
$ \beta\Delta E $ given only $\Delta X_{i,l}$ and the  
field values at the surrounding spacetime points as input features.  
Here, we include nearest- and next-nearest neighbor 
phonon fields in both space and imaginary time, and neglect long-range correlations. While there is justification for a short-range effective interaction  
in proximity to the CDW phase at half-filling~\cite{DeePRB2019}, 
this approximation can also be systematically improved by taking more input features. 
We have found, however, that next-nearest-neighbor inputs are sufficient. 
We also supply an additional 
neuron in the input layer that enforces known behavior at large displacements \cite{Chen2018PRB,Supplement}. 

\begin{figure}[t]
\center\includegraphics[width=\columnwidth]{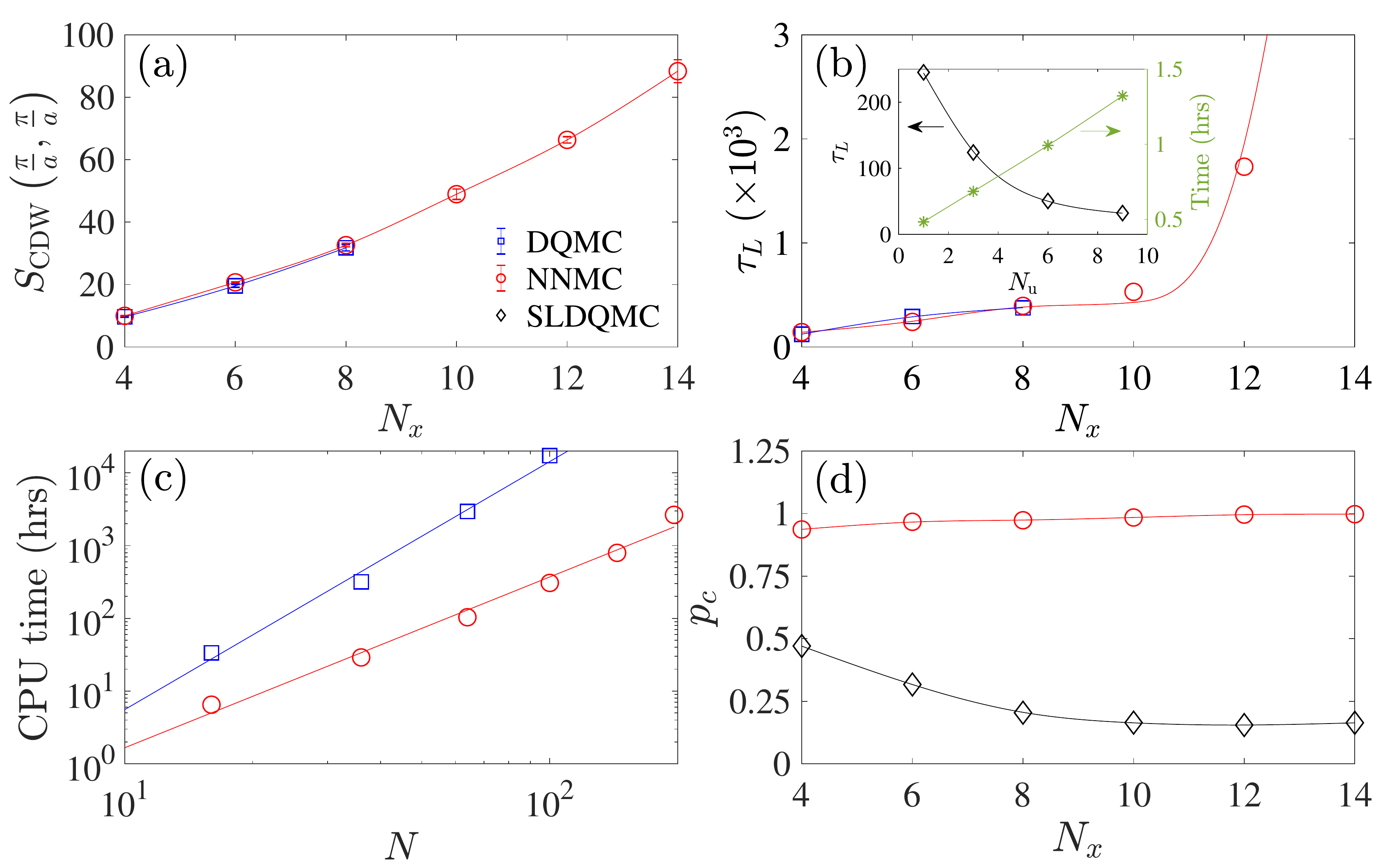}
\caption{\label{Fig:Autocorrelation} 
(a) The CDW structure factor
$S_\text{CDW}\left(\frac{\pi}{a},\frac{\pi}{a}\right)$ and (b) its autocorrelation time as a function of the 
linear cluster size $N_x$ obtained with conventional DQMC and NNMC algorithms. The inset shows the 
reduction of the autocorrelation time and increase of the simulation runtime as the 
number of update sweeps per Monte Carlo sweep $N_\mathrm{u}$ is increased. 
(c) A comparison of CPU time to complete $8\times 10^4$ warm-up and 
$8\times 10^{4}$ measurement sweeps as a function of $N_x$ using the conventional and neural network 
sampling schemes. In both cases, we performed global updates randomly at all sites in the cluster after every one full spacetime sweep of local updates. To make a robust comparison between 
the two methods, we took identical parameters for both sets of simulations. 
The solid lines are fits to the data of the form $t_\mathrm{CPU} = \alpha N^z$.  
(d) The cumulative update ratio of the NNMC algorithm compared against the values 
achieved using the SLMC method as described in Ref. \onlinecite{Chen2018PRB}.    }
\end{figure}

{\it Global Updates} --- Global updates are performed using a convolutional neural network (CNN) 
with four layers [Fig.~\ref{Fig:NN}(b)], where the objective again is to predict $\beta \Delta E$ given only local 
information about the phonon fields. Assuming the update occurs at site $i$, the input layer 
has three columns of input features: the first contains fields
$X_{i,l}$ across all imaginary time slices; the second and third columns contain 
averages $\bar{X}_{i,l}^{(1)}=\frac{1}{4}\sum_{\langle j \rangle}
X_{j,l}$ and $\bar{X}^{(2)}_{i,l}=\frac{1}{4}\sum_{\langle\langle j \rangle\rangle}
X_{j,l}$, respectively, at all time slices,  
where $\langle j \rangle$ and $\langle\langle j \rangle\rangle$ 
denote nearest- and next-nearest-neighbor sums around site $i$. 
The use of $\bar{X}_{i,l}^{(1)}$ and $\bar{X}^{(2)}_{i,l}$ enforces $C_4$ rotational 
symmetry and reduces the cost of training the CNN.  
The convolution operation from the input layer to the first hidden layer is standard \cite{Supplement}.

\begin{figure}[t]
\center\includegraphics[width=\columnwidth]{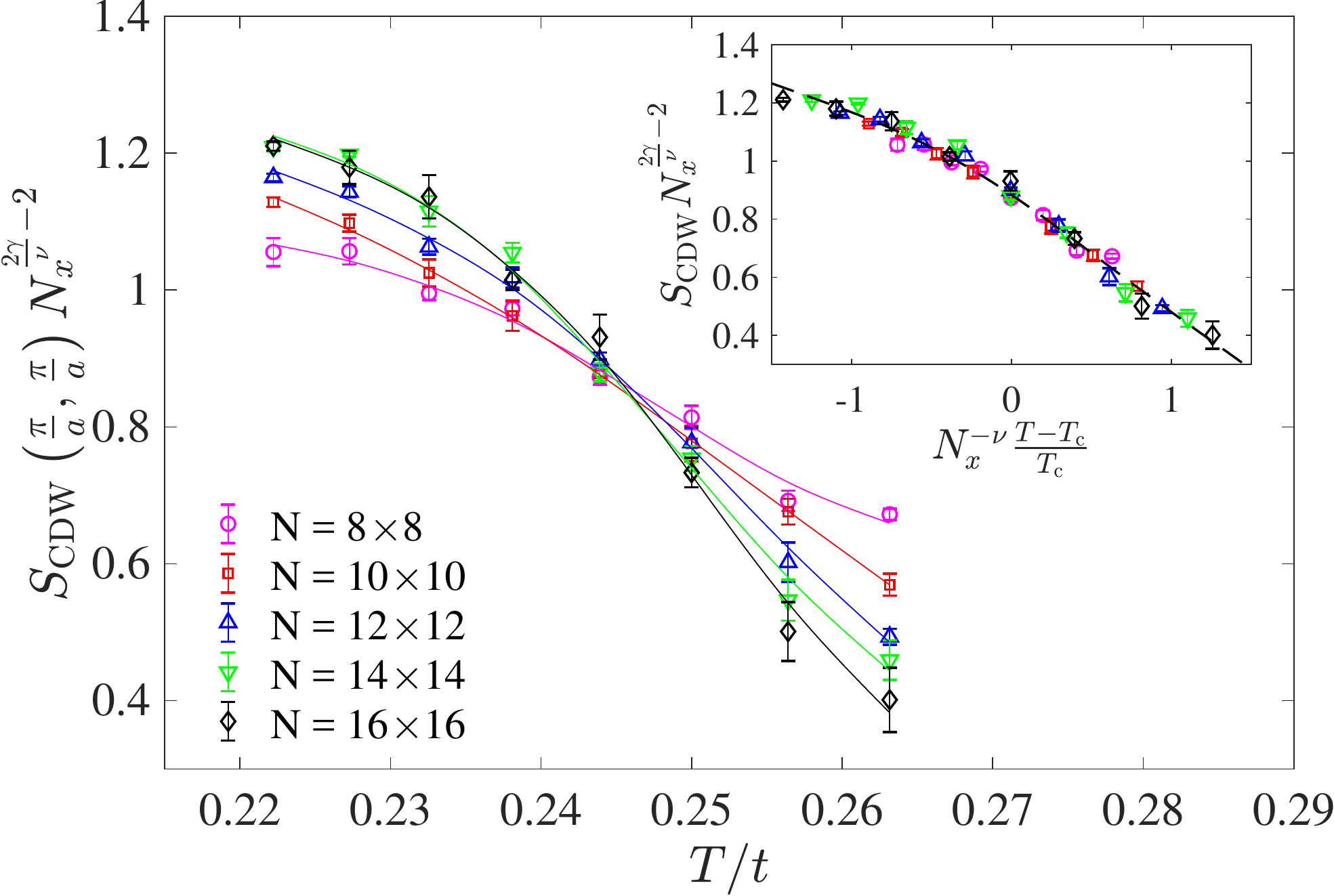}
\caption{\label{Fig:Results} A finite size scaling analysis of the ${\bf q} = \left(\pi,\pi\right)/a$ 
CDW structure factor 
$S_\text{CDW}\left({\bf q}\right)N_x^{2\gamma/\nu-2}$ {\it vs}. $T/t$.  
The CDW transition is in the 2D Ising universality class with 
critical exponents $\gamma = 1/8$ and $\nu = 1$.  
The inset shows the collapse of the data when $S_\text{CDW}({\bf q})N_x^{2\gamma/\nu-2}$
is plotted {\it vs}. $N_x^{-\nu}\frac{T-T_\text{c}}{T_\text{c}}$ with a critical 
temperature $T_\text{c}=0.244$.}
\end{figure}

For each set of (fixed) input parameters $ \{\beta,\mu,\Omega,g\}$ 
we train both networks using training examples generated with 
the conventional DQMC algorithm on a $6\times6$ cluster. 
Throughout, we generated $8\times 10^4$ samples, which were 
randomly partitioned into $6\times 10^4$ training and $2\times 10^4$ test samples. 
We first show results for their performance; the insets of Figs. \ref{Fig:NN}(a) and \ref{Fig:NN}(b) 
compare the predicted $\beta \Delta E_\text{NN}$ against the exact $\beta \Delta E$ values 
obtained from our test data sets for the local and global updates, respectively. 
This simulation was performed close to the CDW transition for the model [Fig.~\ref{Fig:Results}]. 
Both networks accurately predict the MC configure weights but the 
fully-connected neural network is slightly more accurate. While the accuracy can be systematically 
improved by taking more input features, we find that the knowledge learned by both networks can be 
transferred to larger clusters remarkably well based on Fig. \ref{Fig:Autocorrelation} (a). 

Once our networks have been trained and tested, we then define a full MC sweep as consisting of $N_\text{u}$ 
complete sweeps of local updates performed at each spacetime point ($i,l$) using the 
fully-connected neural network, followed by $N_\text{u}$ sweeps of 
global updates performed at {\it every} lattice site $i$ using the CNN. (This sampling procedure differs from the conventional 
one \cite{JohnstonPRB2013}, where global updates are performed on a subset of sites to minimize 
the total computational cost.) After performing these sweeps, the original 
field configuration  $\{X\}$ is replaced with a newly proposed one $\{X^\prime\}$ in a cumulative 
update \cite{LiuPRB2017, LiuPRB2,XuPRB2017} with a probability $\mathrm{min}\left[ 1,p_c\right]$, 
where   
\begin{equation*}
p_c=\frac{W(\{X^\prime\})}{W(\{X\})} 
\frac{\mathrm{exp}(-\beta E_\text{NN}[\{X\}])}{\mathrm{exp}(-\beta E_\text{NN}[\{X^\prime\}])}.  
\end{equation*}

{\it Benchmarks} --- To benchmark the NNMC, we performed direct comparisons 
with the conventional DQMC algorithm for the half-filled model $\langle \hat{n}\rangle = 1$ 
at $\beta=4.1/t$, which is close to the CDW transition
temperature for this parameter set. 
We emphasize that both the DQMC and NNMC simulations used the same sampling protocol  
with  $N_\text{u}=1$. 
Figure~\ref{Fig:Autocorrelation}(a) plots the CDW  
structure factor $S({\bf q})$ at ${\bf q} = (\pi,\pi)/a$ \cite{Supplement} as a function of the
linear cluster size $N_x$, and demonstrates that the NNMC algorithm accurately 
reproduces the results of the conventional DQMC algorithm for the accessible lattice sizes.  
Figure~\ref{Fig:Autocorrelation}(b) compares the autocorrelation 
time $\tau_L$ of $S_\text{CDW}({\bf q})$ for both techniques, which again yields similar
results. We note, however, that the autocorrelation time can be reduced significantly by  
increasing the number of update sweeps $N_\text{u}$ that are performed before computing the 
cumulative update acceptance probability, as shown in the inset of Fig.~\ref{Fig:Autocorrelation}(b).  

To address how NNMC reduces the computational cost, we compare the time to 
solution for both algorithms in Fig.~\ref{Fig:Autocorrelation}(c). 
Fitting a power law $t_{\text{CPU}}=\alpha N^z$ to the data yields 
$z = 3.41$ and $2.35$ for DQMC and NNMC, respectively, 
a significant reduction in the scaling. We note that a similar speedup was 
obtained using SLMC~\cite{Chen2018PRB}; however, the NNMC does not 
require the functional form of the effective model to be specified {\it a priori}. 
Moreover, the NNMC method is more efficient at generating accepted MC  
moves, particularly as it is generalized to larger system sizes. We highlight
this aspect in Fig~\ref{Fig:Autocorrelation}(d), which shows the cumulative acceptance ratio
$p_c$ obtained using NNMC and compares it with SLMC.   
As the methods are generalized to larger cluster sizes, 
$p_c$ decreases for the SLMC method while 
the NNMC method proposes cumulative moves that are almost always 
accepted, and becomes more accurate on larger cluster sizes. 
The decrease of $p_c$ in SLMC is due to the poor performance of the regression model for 
predicting global updates, which requires a more sophisticated effective model.  

We now demonstrate that the NNMC approach can also be used to study the finite-size  
scaling of the CDW structure factor and obtain the transition temperature 
in the thermodynamic limit. Fig. \ref{Fig:Results} presents a 
similar analysis carried out in the same temperature region $\beta=3.8/t \sim 4.5/t$ 
using the NNMC approach. At the critical point, the finite-size scaling behavior has the form 
$S_\text{CDW}(\pi,\pi)/N_x^2=N_x^{-2\gamma/\nu}f\left(N_x^{1/\nu}\frac{T-T_\text{c}}{T_\text{c}}\right)$, 
where $\gamma=\frac{1}{8}$ and $\nu=1$ are the 2D Ising critical exponents. 
The critical temperature  $T_\text{c}/t\approx 0.244$ ($\beta_c=4.1/t$) is determined by the common intersection point 
of the curves. The inset of Fig.~\ref{Fig:Results} replots $S_\text{CDW}N_x^{-7/4}$ 
against $N_x^{1/\nu}\frac{T-T_\text{c}}{T_\text{c}}$, showing the expected data collapsing to a single curve, 
consistent with Ref.~\cite{Chen2018PRB}.

{\it Summary and Conclusions} --- We have extended the use of artificial neural networks 
in self-learning Monte Carlo methods to lattice Monte Carlo simulations. Our approach 
overcomes many of the scaling issues associated with other SLMC implementations and can be widely 
applied to classical and quantum Monte Carlo simulations on extended lattices. We then applied this 
methodology to DQMC studies of the Holstein model. In doing so, we designed 
fully-connected and convolutional neural networks capable of performing accurate {\it local} and, for the 
first time, {\it global} moves in configuration space. Using this method we are able to reproduce 
results of the charge-density-wave transition in this model but using an approach 
that does not require the form of the effective model to be specified in advance. 

Our approach constitutes a generalizable method for performing machine-learning accelerated lattice Monte Carlo simulations (even the Fermi-Hubbard model), provided that the neural networks are sophisticated enough to learn the underlying effective model for the relevant parameter ranges.  In this context, we note that the specific network architectures depicted in Fig. \ref{Fig:NN} accurately predict the CDW transition; however, they are not well-suited to predict the metal-to-superconducting transition away from half-filling. This is also true for the effective model used in the SLMC approach as applied to the Holstein model in Ref. \cite{Chen2018PRB}. To learn and predict $\beta\Delta E$ near the superconducting transition, different network architectures that can capture the physics in this region should be considered. Finally, we stress that our approach allows to compute physical quantities that can be compared with experiments. Previous quantum Monte Carlo + machine learning approaches used to study the Hubbard model on extended lattices \cite{k_chng_17} have generally focused on classifying phases in the parameter space, while physical measurable quantities (such as one- and two-particle functions) are inaccessible.

{\it Acknowledgements} ---  
S. L., P. D., and S. J. are supported by the Scientific Discovery through
Advanced Computing (SciDAC) program funded by the U.S. Department of Energy, Office
of Science, Advanced Scientific Computing Research and Basic Energy Sciences,
Division of Materials Sciences and Engineering.
S. L. is also supported by the National Science Foundation (NSF) under Grant No. DMR-1606348.
E. K. is supported by the National Science Foundation Grant No. DMR-1609560. 
An award of computer time was provided by the INCITE program. This research
also used resources of the Oak Ridge Leadership Computing Facility, which is a
DOE Office of Science User Facility supported under Contract DE-AC05-00OR22725, and the Spartan high-performance facility at San Jose State University, which is supported by the NSF under Grant No. OAC-1626645.

\bibliography{reference}
\end{document}